\documentclass[12pt]{iopart}
\usepackage{iopams}
\usepackage{graphicx}
\usepackage{cite}

\newcommand{\degrees}[0]{\ensuremath{^{\circ}\,}}
\begin{document}

\title{Low Cost and Compact Quantum Key Distribution}

\author{J L Duligall$^{1}$,
M S Godfrey$^{1}$,
K A Harrison$^{2}$,
W J Munro$^{2}$
and J G Rarity$^{1}$}

\address{$^{1}$ Department of Electrical and Electronic
Engineering, University of Bristol, University Walk, Bristol, BS8 1TR}
\address{$^{2}$ Hewlett-Packard Laboratories,
Filton Road, Stoke Gifford, Bristol, BS34 8QZ}
\ead{joanna.duligall@bristol.ac.uk}

\begin{abstract}
We present the design of a novel free-space quantum cryptography
system, complete with purpose-built software, that can operate in
daylight conditions.  The transmitter and receiver modules are built
using inexpensive off-the-shelf components. Both modules are compact
allowing the generation of renewed shared secrets on demand over a
short range of a few metres. An analysis of the software is shown as
well as results of error rates and therefore shared secret yields at
varying background light levels. As the system is designed to
eventually work in short-range consumer applications, we also
present a use scenario where the consumer can regularly `top up' a
store of secrets for use in a variety of one-time-pad and
authentication protocols.
\end{abstract}

\maketitle

\section{\label{Intro}Introduction}

Quantum cryptography provides a means for two parties to securely
generate shared secret material. In practice, this means that two
parties can amplify an initial store of shared secrets. This shared
secret may be used in three primary ways: to protect the Quantum Key
Distribution (QKD) algorithm itself in order to generate new shared
secrets, to identify themselves to each other, and to act as an
encryption key to classically encrypt messages being sent between
themselves. It is usual to delete the shared secret once it has been
used. Consequently, the shared secret should be thought of as being
consumable. Secrecy of the shared secret generation is safeguarded
by encoding information on non-orthogonal quantum states which an
eavesdropper cannot measure without disturbing. Quantum key
distribution protocols are designed in such a way as to detect these
disturbances and thus alert the two legitimate users to an
eavesdropper's presence. Whilst QKD, in principle, is provably
secure against an attacker using technology, both realistic and
theoretical, experimentalists in the field are set the challenge of
developing a system using current methods and apparatus whilst
maintaining this ideal.

In this work we present a low cost QKD system that is aimed at
protecting consumer transactions. We are willing to compromise a
little on performance while retaining the high security associated
with quantum protocols. The design philosophy is based on a future
hand-held `electronic credit/debit card' which communicates with
consumer outlets (an Automated Teller Machine (ATM), for example)
using free space optics.  This device then also acts as a store of secrets shared only with
the bank (or central secure server) which can be used to protect
online transactions. With quantum key distribution protecting the interface between the ATM and the users handheld device, there is no possibility of an eavesdropper gaining key information via `skimming' attacks whereby the key and card details are read using a so-called `false front' on the ATM itself.

\begin{figure}[ht]
    \centering
    \includegraphics[width=250pt]{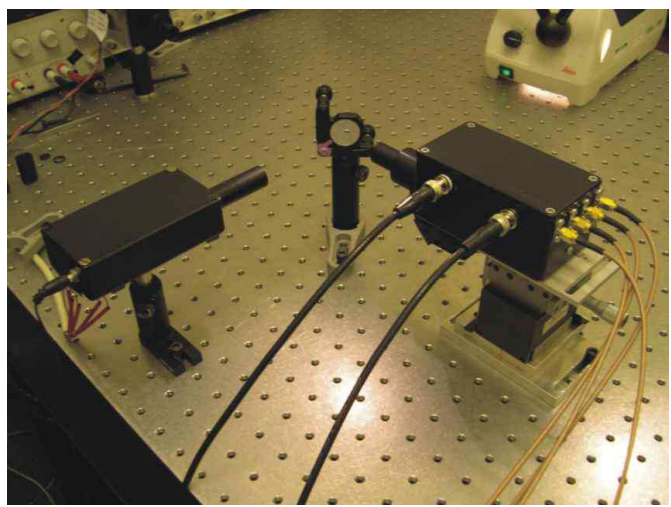}\\
    \caption{The quantum cryptography kit.}\label{qckit}
\end{figure}

This paper briefly introduces the most commonly used quantum
cryptography protocol devised by Bennett and Brassard in
1984\cite{Bennett84} since it has been implemented in the system
presented.  We give an overview of current research and aims in the
field and then go on to outline our intended short range
application.  Section~\ref{System} describes the proposed system in
detail focusing on how costs were brought down without compromising
overly on performance.  Section~\ref{expt} details the experimental
setup with emphasis on the purpose-built software and then presents
key exchange results under various background light conditions and a
security analysis of the system as a whole.  Finally, the paper
concludes with a discussion of the improvements needed in order to
implement the proposed system within its intended application.

\subsection{\label{BB84}The BB84 Protocol}

The BB84 protocol utilizes the quantum property that orthogonal
polarization states can be fully discriminated and thus can be used
to encode information whilst non-orthogonal polarization states
cannot since measuring one necessarily randomizes the other.  The
protocol encodes information in the rectilinear basis (horizontal
and vertical polarization) and the diagonal basis (45\degrees and
135\degrees polarization). The process begins with the transmitter
(Alice) sending a random sequence of photons polarized in each of
the four states (0\degrees, 45\degrees, 90\degrees and 135\degrees).
The receiver (Bob) performs polarization measurements on the
arriving photons, choosing to measure in the rectilinear or diagonal
basis randomly for each photon.  Alice will not know in what basis
Bob measured the photons and similarly Bob is unaware of Alice's
encoding basis. Once the quantum transmission is concluded, Bob
announces publicly, over a classical communication channel, which
photons he received and in what basis he measured them but not the
actual results. Alice then replies with the instances where Bob
chose the correct basis and they both discard all others.  Alice
will also discard all photons that Bob's detectors did not receive.
If Alice and Bob use a coding scheme of 0\degrees and 45\degrees
representing bit value 0, and 90\degrees and 135\degrees
representing bit value 1, then the random bit string generated is
the raw key.  Alice and Bob's results should theoretically now be
correlated unless eavesdropping has taken place on the quantum
channel. The eavesdropper is required to measure photons in a random
basis uncorrelated to that of Alice and then will reinject photons
with errors. Eavesdropping is thus actively monitored by regularly
measuring the error rate and discarding data where the error rate
exceeds a certain threshold (typically $\sim11\%$)\cite{Shor2000}.

In an experimental setup, errors will exist in the raw key because of a
number of causes, including optical imperfections, background counts
and detector noise.  A process of error correction is therefore
required and a variety of techniques are now being
employed\cite{Brassard1994,Yamamura2001,Buttler2003,Pearson2004}. To
minimise any information that Eve might have gained during the
quantum transmission and, indeed, in the error correcting process, a
further technique known as privacy amplification\cite{Bennett88} is
performed resulting in a secret key shared only by Alice and Bob.  It is also usual to have both message integrity and sender authentication
on all communications over the classical channel in order to defeat
a man-in-the-middle attack by Eve.

\subsection{\label{Trends}Current trends in quantum cryptography}

Research in this area is focussed on several key factors.  The most
obvious areas for improvement are transmission range and rate.  Long
distance free-space QKD experiments have developed to the extent
where a 23km key exchange at night was carried out by Kurtsiefer
\textit{et al}\cite{Kurtsiefer02} in 2002.  Significant progress has been made in daylight QKD operation, initially by Jacobs and Franson in 1996\cite{Jacobs96} and later by Buttler \textit{et al}\cite{Buttler1998a} in 1998.  The current record is a 10km link achieved by Hughes \textit{et al} in 2002\cite{Hughes02}. A study has also taken place suggesting that there are no
technical obstacles in developing a quantum key distribution system
between ground and low earth orbit satellites\cite{Rarity02}.
Systems using optical fibre as the transmission medium have achieved
greater distances with Gobby \textit{et al}\cite{Gobby2004}
achieving key exchange over 122km.  As far as transmission rate is
concerned, several systems are now operating at giga-hertz clock
rates\cite{Gordon2005,Bienfang2004}.

Improving current technologies is imperative for quantum
cryptography's future.  Producing detectors with greater efficiency
as well as moving away from faint pulsed lasers as approximations to
single photons will eradicate many security worries. Improvements in
the reliability and efficiency of true single photon sources are
being made with encouraging results\cite{Alleaume2004}. QKD systems are now being sold as commercial products.  MagiQ
Technologies \footnote{www.magiqtech.com},
IDQuantique\footnote{www.idquantique.com} and
SmartQuantum\footnote{www.smartquantum.com} all offer fibre-based
systems for sale whilst other organisations such as QinetiQ, Toshiba
and NIST have quantum cryptography capabilities. The systems
currently available are expensive, use purpose-built components and
are made-to-order. Moreover, their market base is primarily
organisations such as the military, financial services or high
intellectual property establishments. The work presented in this
paper is a clear departure from this goal. It represents a ground-up
approach to quantum cryptography, exploring the possibilities of
bringing secure electronic communication and data exchange to the
consumer.  This research forms part of the SECOQC network\footnote{www.secoqc.net} and concentrates on that final link in the chain from network to consumer/end user by providing end-to-end security for the user as well as the channel.

\subsection{\label{App}The Application}

Figures of credit card fraud loss in the UK for 2004-2005\footnote{www.apacs.org.uk}
show that the only type of fraud to increase this past year was so
called `CARD-NOT-PRESENT' crime typical of mail order or online
transactions. Here we propose a method of protecting these
transactions using the shared secret stored in a personal handheld
transmitter which is regularly topped up by secure key exchange with
a stationary receiver unit.  The Alice module would be incorporated
within a small device such as a mobile phone, or PDA, and the Bob
module within a large fixed device such as a bank ATM, known as the
``Quantum ATM".

A typical usage scenario would be for a customer to register for
this service in her bank. Once the customer is verified as a
legitimate account holder, she is given a SIM card (or an equivalent
storage device) containing an initial unique secret bit string which
she shares with a central secure server that all ATMs have access
to. This One-Time-Pad (OTP) is then used to authenticate and encrypt
future transactions, whether it be withdrawing cash from an ATM or
buying products online. As each security operation requires the
consumption of some of the shared secret, the user would
periodically revisit the ``Quantum ATM" and `top up' both their copy
and the Bank's copy of the shared secret.

\section{\label{System}The System}

\begin{figure}[ht]
    \centering
    \includegraphics[width=445pt]{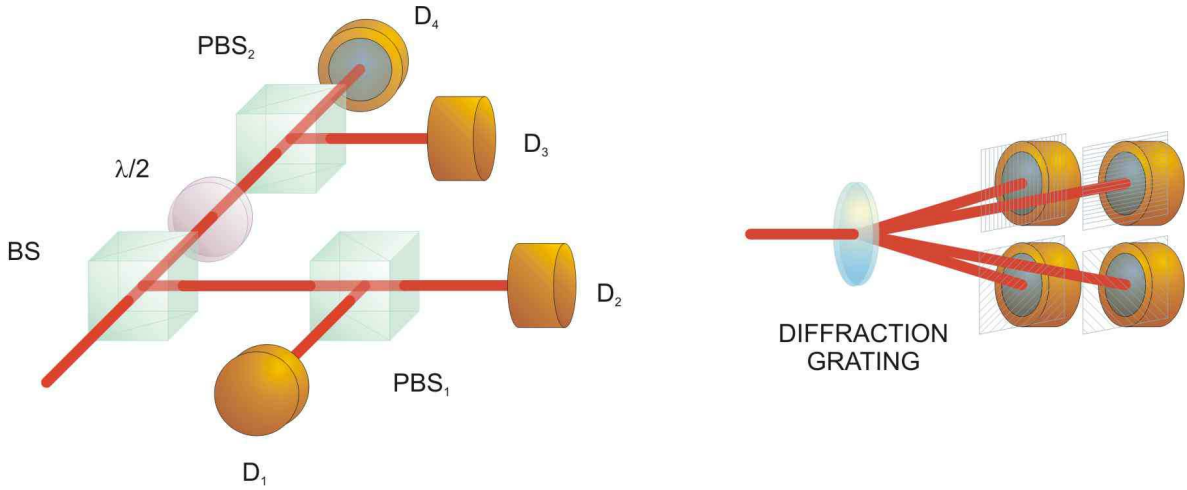}\\
    \caption{On the left, a common implementation of the BB84 protocol using
a beamsplitter cube ($BS$),
    a half-wave plate ($\lambda/2$) and two polarizing beamsplitter cubes
($PBS_{1}$ and $PBS_{2}$).
    The diagram on the right shows a compact system using a
    diffraction grating and dichroic sheet polarizer over each
detector.}\label{cubegrating}
\end{figure}

The system described here is based on the BB84 protocol with a
slight variation from the usual free space experimental
implementation.  Figure~\ref{cubegrating} shows a common optical
arrangement in a receiver unit where the random basis selection and
polarization measurement is made.  For example, a vertically
polarized photon, emitted from Alice is directed through the 50:50
beamsplitter $BS$, making the basis selection. If it is reflected,
the photon is to be measured in the rectilinear basis and the
polarizing beamsplitter $PBS_{1}$ directs the photon to a detector
according to its polarization. Thus $D_{1}$ receives a click.  If
the photon is transmitted at $BS$, the diagonal basis is chosen and
the photon passes through the $\lambda/2$ plate and then $PBS_{2}$
where, in the case of a vertically polarized photon, either $D_{3}$
or $D_{4}$ might click with equal probability.  This result would be
discarded later in the sifting process as Bob measured the photon in
the wrong basis.  Therefore, in general, the BB84 implementation has
a 50\% protocol efficiency as half the photons will be measured in
the wrong basis.  We use a different optical arrangement which acts
to reduce the size of the module as well as the cost.  It involves
the use of a holographic diffraction grating to produce a 2x2 matrix
of beam paths (figure~\ref{cubegrating}).  The grating therefore
makes the random basis selection by sending an incident photon in
one of four directions.  In order to make the polarization
measurements in accordance with the BB84 protocol, dichroic sheet
polarizer was placed in front of each detector in one of the four
polarization orientations.  Note that in this arrangement, the
protocol efficiency has dropped to 25\% since the photon is directed
randomly in one of four ways.  This trade-off in efficiency vs. cost
was deemed acceptable since there are far fewer transmission losses
in a short range system.

\subsection{\label{Alice}The Alice Module}

\begin{figure}[ht]
    \centering
    \includegraphics[width=445pt]{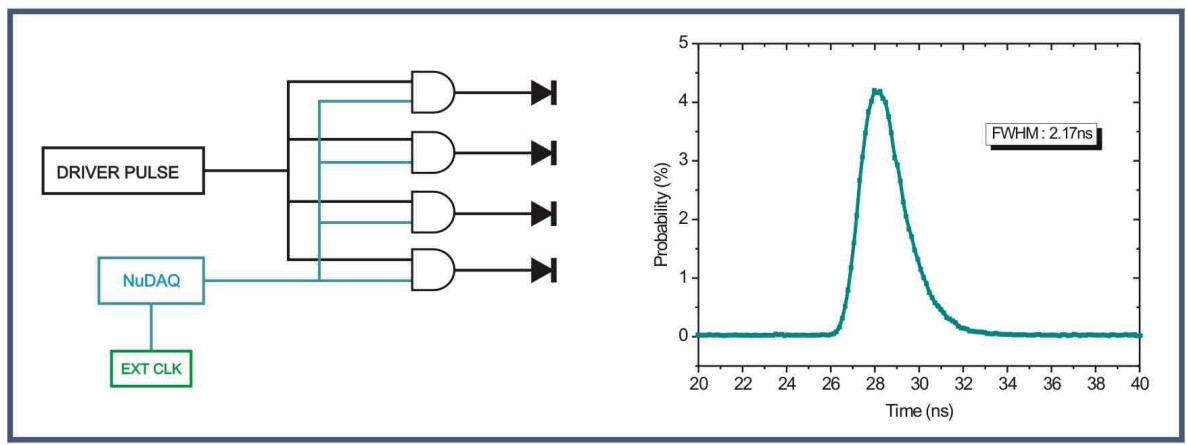}\\
    \caption{The sub-5ns driver pulse is ANDed with the signal from the
digital I/O card (NuDAQ)
    that determines which LED will fire.  The time interval histogram gives
a measure of the optical
    pulse width of one LED from the Alice module.}\label{driver}
\end{figure}

In its experimental form, the Alice module uses off-the-shelf IC
components in a driver circuit which produces sub-5ns pulses.  The
driver pulses are then ANDed with the output from a digital
input/output card (NuDAQ, \textit{Adlink} PCI-7300A) and passed to
one of four AlInGaP, miniature, red-orange LEDs (\textit{Agilent},
HLMA-QH00), see figure~\ref{driver}.  The output from the NuDAQ card
is regulated by an external oven-stabilized clock (\textit{C-MAC
Frequency Products}, CFPO-6) and passes a random bit string,
generated by a quantum random number generator (QRNG),
(\textit{IDQuantique}, Quantis) to the Alice module, recording which
LED fires.  Due to the limitations of the i/o card, the driver
pulses are produced at a repetition rate of 5MHz.  The four LEDs
were intensity balanced by adjusting the current to each diode and
timing jitter measurements were carried out for each channel
showing, on average, pulses of $2.4 ns$ duration.
Figure~\ref{driver} also shows the resulting time interval histogram
of one channel.

\begin{figure}[ht]
    \centering
    \includegraphics[width=250pt]{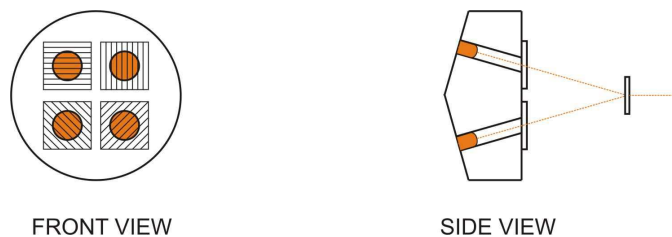}\\
    \caption{Front and side view of the holder designed to house the LEDs.
    The front view shows the dichroic sheet polarizer placed over each LED
output.}\label{holder}
\end{figure}

The LEDs are attached to a holder, see figure~\ref{holder}, with
dichroic sheet polarizer, orientated in each of the four
polarization states, 0\degrees, 45\degrees, 90\degrees and
135\degrees, placed over each output. To combine the beam paths, the
same four way diffraction grating arrangement was used.  The holder
serves to direct the polarized light towards the grating as well as
restrict the viewing angle of the diodes. A pinhole was placed after
the grating together with a 50mm focal length lens to collimate the
beam. A $632.8\pm3 nm$ filter was included to limit the bandwidth.

Alice and Bob communicate via the internet. This is assumed to be a
public channel. In the intended use model, this will be replaced by
an IrDA infra red communication channel.

\subsection{\label{Bob}The Bob Module}

In its experimental form, the Bob module contains four passively
quenched silicon avalanche photodiodes (\textit{Perkin Elmer},
C30902S) which are cooled down to -10\degrees C and maintained by a
temperature controlling circuit.  Since these detectors operate at a
relatively high voltage, a high voltage DC-DC converter
(\textit{EMCO}, Q03-5) was included so that the Bob module can run
off a low voltage supply.  A simple discriminator circuit takes the
output from the detectors and converts it to a readable positive
pulse.  Time of arrival information is recorded by a time interval
analyser card (TIA, \textit{GuideTech}, GT653).

In a similar setup to the Alice module, the dichroic sheet polarizer
was placed in front of each detector orientated in the four
polarization directions with the diffraction grating in place, as
shown in figure~\ref{cubegrating}. In addition to this arrangement,
a $632.8\pm3 nm$ filter was included to reduce the background count
and a 50mm focal length lens to collect the beam from Alice and
focus it down onto the detectors.
\\
\\
The bit error rate (BER) for each channel was estimated from data
taken during key exchange (see section~\ref{expt}).

\begin{eqnarray}
BER=\frac{N_{\textit{wrong}}}{N_{\textit{total}}}
\label{BERcalc}
\end{eqnarray}
\\
where $N_{wrong}$ is the number of bits in error and $N_{total}$ is
the number of bits received in total.
\\

This gives a measure of the likelihood of Bob receiving a 0 when a 1
was sent from Alice.  All but one of the BER values in
table~\ref{Table:BER} are sufficiently low showing that optical
imperfections from the equipment will contribute little to the error
in the sifted key.  Unfortunately, the over-sensitivity of one
detector in the Bob module is responsible for the greater BER in the
135\degrees channel.  Whilst this problem does increase the overall
base error rate of the system and ultimately bias Bob's data opening it up to possible attacks\cite{Makarov05}, this
was deemed acceptable with the current set of equipment providing
proof-of-principle results. An updated version of the receiver
module is in development with balanced detectors and a more
efficient cooling system to lower the dark count rates.

\begin{table}[h]
\caption{\label{Table:BER}Table showing the BER values at 0\degrees,
45\degrees, 90\degrees and
135\degrees as percentages calculated from equation~\ref{BERcalc}.}
\begin{indented}
\item[]\begin{tabular}{@{}ll}
\br
Channel&Bit Error Rate (\%)\\
\mr
0\degrees & $1.32$\\

45\degrees & $2.54$\\

90\degrees & $2.20$\\

135\degrees & $4.75$\\
\br
\end{tabular}
\end{indented}
\end{table}

\subsection{\label{Loss}Loss Tolerance of a Daylight System}

Since the module will have to be able to operate in daylight
conditions, the background error rate\cite{Rarity02} is the most
important consideration in designing the receiver unit. It will be
the limiting factor of the entire system.

The signal count for this system is defined as

\begin{eqnarray}
S=\frac{RMT\eta}{4}
\label{eqn:signal}
\end{eqnarray}
where $R$ is the pulse repetition rate, $M$ is the average number of
photons per pulse, $T$ is the lumped transmission (including
geometric loss) and $\eta$ is the detection system efficiency. The
protocol effectively splits the signal into four on the detectors
leading to a factor of 4 reduction in signal bit rate after sifting
the key.  The background rate is given by

\begin{eqnarray}
P_{b}=Bt
\label{eqn:noise}
\end{eqnarray}
where $B$ is the background count rate per detector and $t$ is the
time synchronization gate. Half these background counts induce an
error and half are thrown away in the protocol but contributions
arise from all four detectors.
\\
Using equations (\ref{eqn:signal}) and (\ref{eqn:noise}), the
background error rate is thus given as

\begin{eqnarray}
E=E_{base} + \frac{P_{b}}{S}
\label{eqn:E}
\end{eqnarray}
A base error rate of $E_{base}=0.027$ (derived from an average of
the values in table~\ref{Table:BER}) is expected thus

\begin{eqnarray}
E=0.027+\frac{4Bt}{MT\eta}
\label{eqn:E1}
\end{eqnarray}
\\
Error correction schemes will operate efficiently with an error rate
of $E<0.08$, therefore the maximum acceptable background count rate
per detector is given as

\begin{eqnarray}
B < \frac{MT\eta}{75.5t} \label{eqn:B}
\end{eqnarray}
\\
In considering the system presented here, estimates can be made for
the following values:

\begin{itemize}
    \item $M \sim 0.3$, an accepted value for guaranteed security of low
loss systems, using the optimal choice for the expected photon number taken from \cite{Lutkenhaus00}.
    \item $T \sim 1$ since the source can be imaged onto the receiver and
the system is short range
    and thus atmospheric loss is negligible.
    \item $\eta \sim 0.045$ taking into account the quantum efficiency of
the detectors and the presence
    of the narrowband filter and polarizers.
    \item $t = 5 ns$ gate synchronization time.
\end{itemize}
Thus the maximum background count rate per detector can be given as
roughly
\\
\\
\centerline{$B\leq 36000$ Counts/sec}
\\
\\
The current version of the Bob module can operate in shaded areas but not in full direct sunlight.  Of course higher error rates are in part due to a relatively wide time synchronization gate.  This is due to the limitation of using LEDs to produce short pulses.  The timing window can be shortened further but then the bit rate is reduced, as is shown in section~\ref{Future}, figure~\ref{Comp}.  An updated version is under development whereby a restriction in the field of view and greater spectral filtering will be introduced.  Shared secret generation has been carried out at background light levels of up to $26000$ counts/sec and the results are shown in section~\ref{expt}.  

\begin{figure}[ht]
    \centering
    \includegraphics[width=350pt]{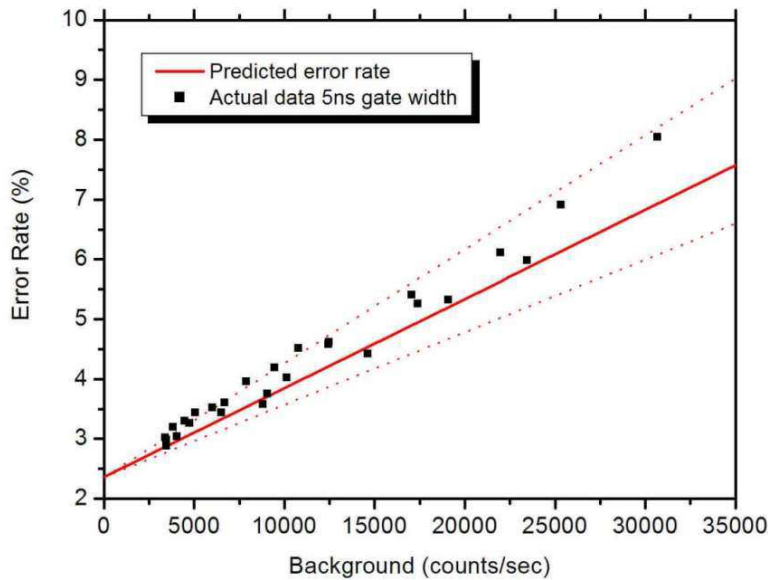}\\
  \caption{Graph showing the relationship between the estimated error rate
and the background
  counts per second  The red line is the predicted error rate calculated
from equation~\ref{eqn:E1}.
  Our current Bob module has detectors of slightly varying detection
efficiency.  The dashed lines
  indicate the predicted error rate with $\eta$ at $0.045$ and
$0.055$.}\label{error}
\end{figure}

In order to ascertain how well the hardware compared to equation~\ref{eqn:E1}, a random data string was sent from Alice to Bob and the percentage of errors at varying backgrounds was calculated.  Figure~\ref{error} shows this relationship as well as the predicted error rate for each background level.  The dashed red lines are essentially an upper and lower bound to the predicted error rate since the channels within the Bob module have slight differences in efficiency.  As is shown, the data points all fit well within these bounds.  Again, an updated version of the Bob module is expected to increase the value of $\eta$ to about $0.08$ per detector and thus raise the background level at which the system can operate to over $60000$ counts per second.  Also note that our two channel measurement system (see section~\ref{software}) effectively doubles the background rate per detector and a four channel measurement system will further improve our resilience to background light.

\section{\label{expt}Experimental Setup}

\begin{figure}[ht]
    \centering
    \includegraphics[width=445pt]{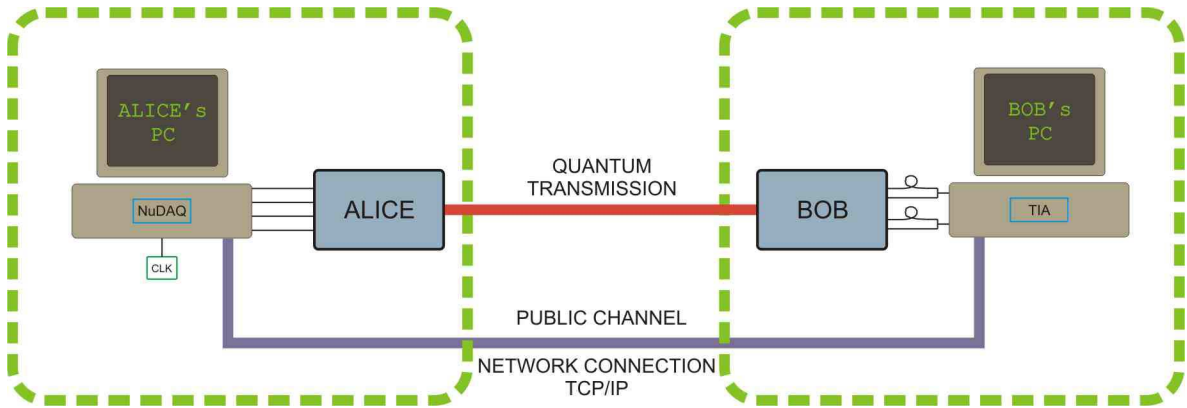}\\
    \caption{Schematic showing the experimental setup for demonstrating key
exchange.}\label{setup}
\end{figure}

The experimental setup for quantum key distribution is shown in
figure~\ref{setup}.  As mentioned in section~\ref{Alice}, a file
containing a random bit string generated from the QRNG is supplied
to the NuDAQ which in turn controls the Alice module at a repetition
rate of 5MHz.  The quantum transmission is received by the Bob
module which passes the four channel outputs to the TIA card.

\subsection{\label{software}Software}

The key requirements from the software is that it should be fast
establishing a key within seconds. This should work with minimal
interactivity between Alice and Bob and with low processing power at
Alice. The synchronization (of both clock and start time) and error
correction software have been developed with these constraints in
mind.

\subsubsection{Synchronization}

In this system, the data is recorded first during the quantum
transmission and processed afterwards in a few seconds.  The start
of the transmission is determined approximately by searching for a
jump in the frequency of time tags as Bob starts measuring before
Alice begins her transmission. Alice transmits sub-$5 ns$ pulses
every $200 ns$, therefore a time synchronization gate of $5 ns$
reduces the probability of registering a background event within the
gate by a factor of 40. The clock at Bob is thus synchronised with
the clock at Alice by searching for time tags that sit at
separations of $200 ns$ and adjusting the time separation slightly
every $\sim100 ms$ to compensate for clock drift. The advantage of this
setup is that no timing reference signal is needed. To determine the
exact start time of the data Bob reveals a random subset of his
measured bit values and the basis he used to Alice. Alice then finds
the data start by performing a sparse correlation against her stored
data. This random subset can also be reused to estimate the error
rate.

\begin{figure}[ht]
    \centering
    \includegraphics[width=445pt]{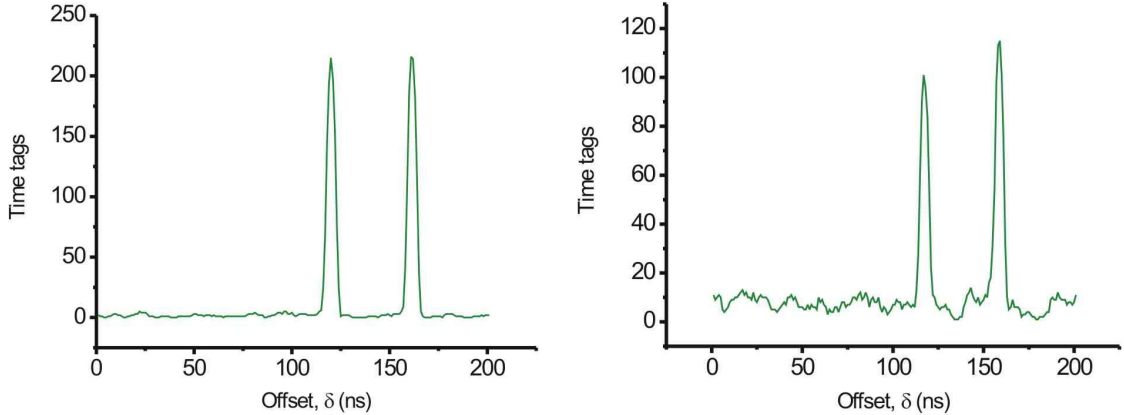}\\
    \caption{Two histograms showing the results of running the
synchronization routine on
    the data from the TIA card.  The card has only two input channels so two
channels
    from Bob are coupled onto the other two with a fixed delay of $40ns$
accounting
    for the two peaks shown.  The right hand graph shows how the noise
increases
    when the experiment is run in higher background light
levels.}\label{gates}
\end{figure}

Since the GT653 is a two-input card, we combine two channels into
one input by delaying one channel by ~$40 ns$, see
figure~\ref{gates}. In doing this, the background count rate is
effectively doubled. This is not currently a problem since we can
determine the true signal to noise ratio from the data and calculate
error rates and thus secret bit yields against this background.
However, the background count rates shown are not per detector but
per channel and so using the TIA card in this manner does become a
limiting factor when operating at higher background light levels as
we are effectively making the situation worse.  Improvements to this
setup are discussed in section~\ref{Future}.

\subsubsection{Error correction}

Most quantum cryptography systems use the Cascade
\cite{Brassard1994} algorithm since it operates close to the
theoretical Shannon limit.  It is highly interactive involving many
separate two-way communication steps. Any latency in the classical
channel dramatically slows the process. Thus another method of error
correction has been adapted for this system.  We have chosen a
version of the Low Density Parity Check (LDPC)\cite{Gallager62}
algorithm as the protocol has very little interactive communication.
In fact, Alice merely needs to transmit an error correction syndrome
to Bob. One drawback is that LDPC requires a pessimistic lower bound
estimate of the error rate. The synchronization protocol already
provides us with an error estimate. Information revealed during the
error correction process is removed by a privacy amplification
process \cite{Bennett88} in which key length is reduced.

It should be noted that our implementation of error correction
requires that Alice and Bob both generate the same random factor
graph. Once Alice and Bob know the number of message bits they are
error correcting over, and the measured error rate, they seed a
pseudo random number generator from their OTP and use this to
generate an appropriate factor graph. Eve, the eavesdropper, is
assumed not to know which of the $2^{256}$, say, different factor
graphs Alice and Bob are using.

\subsection{\label{Results}Experimental Results}

\begin{figure}[ht]
    \centering
    \includegraphics[width=350pt]{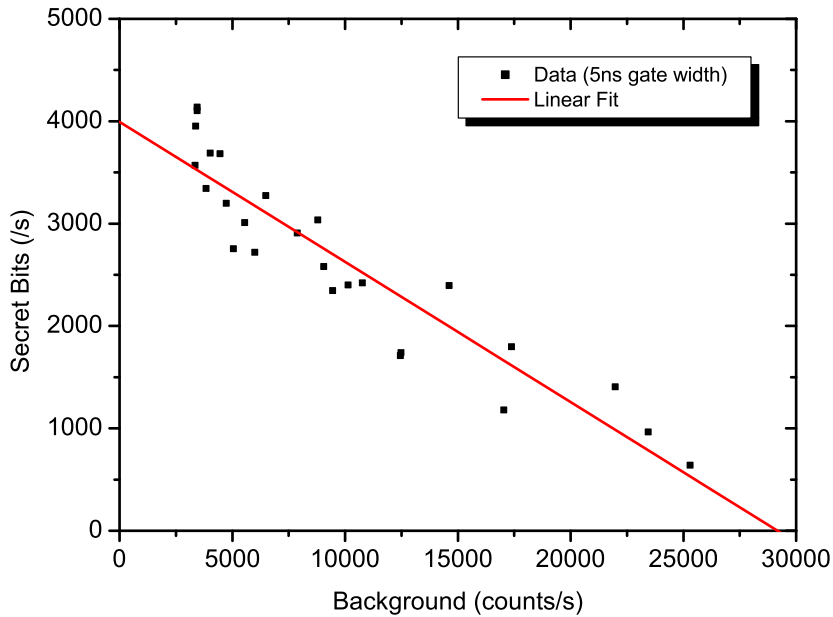}\\
  \caption{Graph showing the relationship between the secret bit rate and
the background
  counts per second.}\label{secret}
\end{figure}

Using the experimental setup described in section~\ref{expt}, we
carried out a series of key exchange runs at varying background
light levels. From the data we extracted an error rate using the methods described above. The error corrected keys were then passed through the privacy amplification process to remove all possible information leaked to a theoretical eavesdropper. This effectively reduces the length of the key but ensures absolute secrecy. To estimate the key length reduction, we use the Lutkenhaus bound\cite{Lutkenhaus99} and details are presented in the appendix. The resulting number of bits divided by the collection time thus gives us an equivalent secret bit rate as a function of background counts which we show in figure~\ref{secret}. We are able to establish just over 4000 secret bits per second at low background with over 500 secret bits at count rates exceeding 25000 counts/sec.  We note here that we have maintained a very pessimistic view of the eavesdropper's capabilities and there may be the possibility of increasing the number of secret bits without incurring too strong a security penalty. Our near term target will be to improve the system to the point where 10000 secret bits can be generated even with backgrounds of order 30000 per second. We can immediately do this if we double the repetition rate to 10MHz, improve the Bob module detection efficiency to $8\%$ and increase the protocol efficiency from $25\%$ to $50\%$.  We discuss this further in the following section.    

\section{\label{Future}The Future}

The initial experimental setup has shown that off-the-shelf
components can be used to carry out quantum key distribution.  We
are currently working to produce stand-alone modules for the
transmitter and receiver. An inexpensive Field Programmable Gate
Array (FPGA) will be used to replace much of what has been achieved by
software on Alice's computer as well as replace the driver circuit
shown in section~\ref{Alice}.  The transmission rate of this system will be increased to 10MHz. We will interface the FPGA Alice module with a PDA (\textit{Hewlett-Packard}, iPAQ hx4700
series) via serial cable and are developing the software to run on the
PDA which will incorporate the IrDA or BlueTooth facilities as the
QKD public channel.  We are also working on an FPGA solution to the
time interval analyser on the receiver side.  A four channel TIA
module with an estimated resolution of $300 ps$ is under
development. 

\begin{figure}[ht]
    \centering
    \includegraphics[width=350pt]{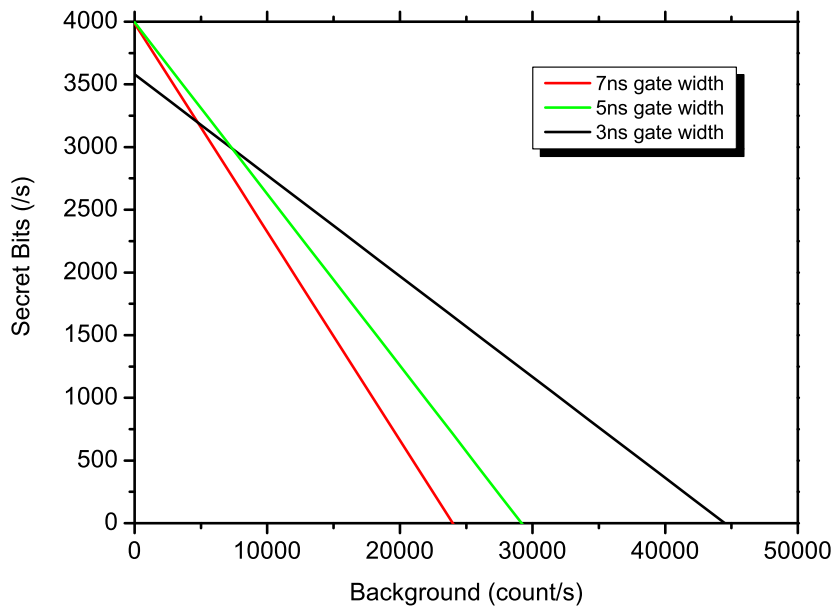}\\
  \caption{Graph showing the relationship between the secret bit rate and
the background
  counts per second using 3,5 and 7ns time synchronization gate
widths.  Since the system must be able to operate in daylight conditions 
there could be a procedure whereby Bob samples the background count 
before synchronization and correlation procedures and chooses which 
gate width to use.}\label{Comp}
\end{figure}

Improvements to the software both in speed and efficiency are under way.  There are plans to dynamically evaluate
Bob's data using various time synchronization gate widths depending on the level of 
background counts. In this way we will still be able to obtain secret bits 
at higher background levels. It is important that we do not waste the OTP stack 
by authentication protocols whilst attempting to carry out key 
exchange in very bright conditions.  Once the OTP is depleted from too many unsuccessful attempts
to extend the key, the user would have to revisit the bank to obtain a new OTP. Figure~\ref{Comp} illustrates how 
a secret bit rate can be achieved at higher background count rates by evaluating Bob's 
data with narrower gate widths. 

On the hardware side, we plan to improve the Bob receiver module.  Initially we will start by using an improved grating and polarizers to increase the efficiency to around $8\%$.  However, as the receiver module can be a static medium cost device, it may prove better to return to a classical beamsplitter based design with protocol efficiency of $50\%$. Eventually the Alice module must be able to be brought to the Bob module and fully automatically aligned. Initially this may be simply done using a fixed alignment cradle or docking station. However, we are also considering active methods for aligning a hand-held Alice module to a stationary Bob module.

\section{\label{Conclusion}Conclusion}

We have built a low cost free-space quantum cryptography
system using off-the-shelf components that is able to generate and renew shared secrets on demand over a
short range of up to a metre in shaded daylight conditions. 
The transmitter unit is compact and we are aiming eventually to incorporate 
it in a hand held device such as a smart card or mobile phone. A full software system has been developed to handle synchronization, error estimation and correction and privacy amplification. 
We have tested the system in a range of background levels up to that equivalent to shaded daylight conditions.
The system is designed to work in short-range consumer applications and we have 
described a use scenario where the consumer can regularly `top up' a
store of secrets for use in a variety of one-time-pad and
authentication protocols. We have described various improvements to the system 
that will increase our background light tolerance and bit rates while reducing cost and complexity.  Currently, our system can generate around 4000 bits of secret keys from a one second interaction between transmitter and receiver in low light conditions. In the next generation device we expect to be able to operate at 10000 secret bits per second up to full daylight conditions.

\section{\label{Acknowledgements}Acknowledgements}
We gladly thank Alastair Lynch 
from the University of Bristol for his continued work on this
project. John Rarity acknowledges support from the Royal Society through a Wolfson Research Merit Award.
Mark Godfrey is funded by EPSRC CASE studentship through QinetiQ.
Joanna Duligall is funded by EPSRC CASE studentship through HP Laboratories.
This work has also been supported by the EU under project number
FP6-2002-IST-1-506813 SECOQC and we would also like to acknowledge
the EPSRC QIP IRC for their support. 
\newpage

\section*{References}
\bibliographystyle{unsrt}
\bibliography{References}
\newpage
\appendix

\section*{Appendix}
\setcounter{section}{1}
The number of sifted key bits received in a transmission of duration
$t$ is given by

\begin{eqnarray}
n_{rec}=St
\label{nrec}
\end{eqnarray}

In using the LDPC coding scheme, the error correction efficiency $E$
is calculated using

\begin{eqnarray}
E=1-n_{syn}/n_{rec}
\label{corrEff}
\end{eqnarray}

where $n_{syn}$ is the number of syndromes needed. Maintaining a
pessimistic view of an eavesdropper's capabilities, we discard all
syndrome bits.  In considering a pulse splitting attack with zero
loss transmission technology\cite{Lutkenhaus99}, the eavesdropper
takes all the pulses at the output of Alice an blocks all single
photon pulses.  Multi-photon pulses are split and sent on through a
loss free channel to Bob.  The split off photons are stored until
the bases are revealed thus making all received pulses insecure. If we assume a Poisson distribution for the faint pulse photon number probability then

\begin{eqnarray}
P(n)=\frac{M^{n}e^{-M}}{n!}
\label{Poisson}
\end{eqnarray}

and

\begin{eqnarray}
P(n\ge2)= 1-e^{-M}-Me^{-M}
\label{p2}
\end{eqnarray}

Secret bits are only gained when the received photon rate $n_{rec}$
is greater than the multi-photon rate. We then calculate the
fraction of received bits that are guaranteed to be secure as

\begin{eqnarray}
b=1-\frac{(1-e^{-M}-Me^{-M})}{(1-e^{-M})T}
\label{b}
\end{eqnarray}

where $T$ is the efficiency of the free space channel.
\\
\\
We cannot use these bits either in the key or to estimate the error rate.
Hence using Lutkenhaus' formulae, the secret bit rate becomes

\begin{eqnarray}
n_{fin}=(n_{rec}-n_{err})b\left(E-log_{2}\left(
1+4\frac{\epsilon}{b}-4\left(\frac{\epsilon}{b}\right)^{2}\right)\right)-n_{
s}
\label{nfin}
\end{eqnarray}
where $n_{err}$ is the number of bits used to estimate the error, $\epsilon$ is the estimated error rate and $n_{s}$ is a safety
margin in this case set to 100 bits.
\\
Calculating $n_{fin}$ for each data point, figure~\ref{secret} shows
how the secret bit rate varies with background count.

\end{document}